\documentclass[9pt,twocolumn,twoside]{opticajnl}
\journal{opticajournal} 
\usepackage{mathrsfs,amsmath}
\usepackage{physics}
\setboolean{shortarticle}{true}

\usepackage{lineno}

\title{Laguerre-Gaussian modes become elegant after an azimuthal phase modulation}

\author[1]{\textbf{Vasilios} Cocotos}
\author[1]{Light Mkhumbuza}
\author[2]{Kayn A. Forbes}
\author[3,1]{Robert de Mello Koch}
\author[1+]{Angela Dudley}
\author[1*]{Isaac Nape}

\affil[1]{School of Physics, University of the Witwatersrand, Private Bag 3, Wits 2050, Johannesburg, South Africa}
\affil[2]{School of Chemistry, University of East Anglia, Norwich Research Park, Norwich, Norfolk, NR4 7TJ, United Kingdom}
\affil[3]{School of Science, Huzhou University, Huzhou 313000, China}

\affil[*]{isaac.nape@wits.ac.za}
\affil[+]{angela.dudley@wits.ac.za}

\begin{abstract}
Laguerre-Gaussian (LG) modes are solutions of the paraxial Helmholtz equation in cylindrical coordinates and are associated with light fields carrying orbital angular momentum (OAM). It is customary to modulate such beams using phase-only vortex profiles, for example, when increasing (laddering up) or decreasing (laddering down) the OAM content of some given LG mode.
However, the resulting beams have been shown to be hypergeometric-Gaussian modes, due to the changing radial amplitudes on propagation. In this work, we show that these beams in fact have the angular spectrum of elegant Laguerre-Gaussian (eLG) modes, and therefore map back to LG-type modes. Accordingly, the fields obtain new OAM and radial quantum numbers that depend on the initial OAM and additional OAM gained during modulation.

\end{abstract}

\setboolean{displaycopyright}{false} 

\begin{document}
\maketitle
The standard Laguerre-Gaussian (LG) beam is well-known as an eigenmode of the paraxial wave equation whose mode order is determined by the topological charge $\ell \in \mathbb{Z}$ and radial index $ p \in \mathbb{N}_0 $. Whilst $\ell$ tells us the number of intertwined helical wavefronts, and consequently the value of OAM $\ell\hbar$ per photon, the radial index determines the number of nodes, or $p+1$ concentric rings of intensity.
Over the past few decades, LG beams have gained significant attention due to their broad applications in areas such as optical trapping and manipulation, classical and quantum optical communications, imaging, metrology, and spectroscopy \cite{shen2019optical, yang2021optical}. Popular passive techniques for generating LG beams include spatial light modulators \cite{forbes2016creation}, Q-plates \cite{rubano2019q}, and metasurfaces \cite{ahmed2022optical}. Active (direct) generation from source can be achieved in free-space, fiber, and on-chip lasers \cite{shen2019optical, forbes2024orbital}.

As extensions of the LG beam, elegant Hermite-Gaussian (eHG) and elegant Laguerre-Gaussian (eLG) beams were introduced in 1973 by Siegman, and in 1985 by Takenaka, respectively, as solutions to the paraxial wave equation for free-space propagation \cite{siegman1973hermite} \cite{takenaka1985propagation}. However, before this, eHG and eLG beams had been explored in the context of resonators with variable-reflectivity mirrors \cite{vakhimov1965open, zucker1970optical} and lens-like gain media \cite{arnaud1971mode, casperson1976beam}. 
Furthermore, eLG modes exhibit more symmetry compared to LG modes \cite{saghafi1998beam, saghafi1998near, zauderer1986complex}. While the LG’s polynomial argument is real, the eLG incorporates a complex argument, which endows them with compactness and elegance, a feature retained even in three-dimensional generalizations (non-paraxial regime) of eLG beams \cite{takenaka1985propagation, zauderer1986complex}. 
Elegant Laguerre-Gaussian beams are complementary to other complex light fields, though due to their complex argument, display distinctive structural changes in their transverse profiles upon propagation that have been studied in detail \cite{yokota1985relations}.
Connections between elegant and standard modes have been established \cite{yokota1985relations}, and eLG beams with high radial orders have been shown to share similarities with Bessel-Gaussian (BG) beams \cite{sheppard1978gaussian, gori1987bessel}, which are finite-energy realizations of the non-diffracting Bessel beam \cite{durnin1987nondiffracting}. Many theoretical studies have also been dedicated to eLG beams pertaining to their vortex structure \cite{martinez2013vortex, nasalski2014vortex}, OAM \cite{lopez2013derivatives, kotlyar2014hermite}, and vector wave packet behavior \cite{nasalski2013exact}. 

In this work, we demonstrate that LG beams undergoing azimuthal phase modulation evolve into \textbf{elegant} LG modes upon propagation. 
It has been previously shown that these types of evolving LG beams can be described as hypergeometric-Gaussian modes
\cite{kotlyar2007hypergeometric, karimi2007hypergeometric, sephton2016revealing} and, when expressed in the complete LG azimuthal and radial basis, the field represents a superposition of LG modes. However, we show here that LG modes undergoing azimuthal phase modulation revert to LG-type fields, specifically the eLG modes characterized by well-defined radial indices and topological charges.
\begin{figure*}[h]
  \centering
\includegraphics[width=0.9\textwidth]{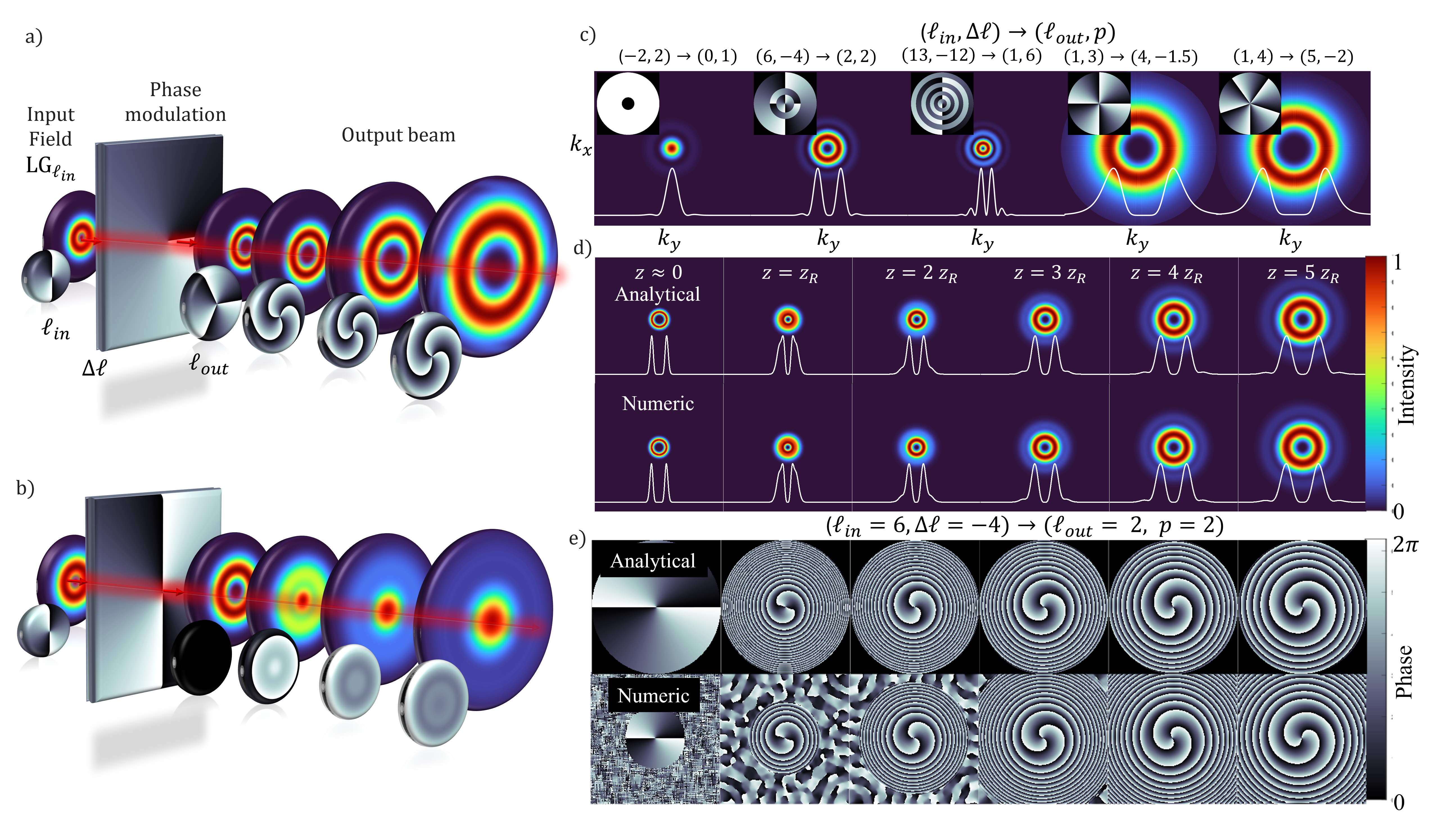}
  \caption[Conceptual figure]{\textbf{Concept figure.} \textbf{a), b)} Conceptual illustration whereby we modulate a given LG mode having topological charge $\ell_{\text{in}}$ with a spiral phase ($\text{e}^{i\Delta \ell \phi}$) corresponding to an OAM charge $\Delta\ell$, thereafter producing a new mode with a topological charge $\ell_{\text{out}} \ = \ell_{\text{in}}+ \Delta\ell $. This is shown analytically,   for (a) $\ell_{\text{in}}= 2$ and $\Delta\ell = 1$ and (b) $\ell_{\text{in}}= -2$ and $\Delta\ell = 2$ over distances $z = 0$ to $z = 1.5z_{R}$ at distance intervals of 0.5$z_{R}$ for both the intensity and phase profiles. \textbf{c)} Analytical  transverse angular spectral profiles for various choices of ($\ell_{\text{in}}$,  $\Delta\ell$) producing OAM topological charges and radial index pairs,$(\ell_{\text{out}}, p)$, with the intensity profiles and their cross-sections in large, and their corresponding phase profiles appearing as insets. The analytical and numerical propagation dependent  \textbf{d)} intensity and  \textbf{e)} phase profiles for the mode created from $\ell_{\text{in}}=6$ and $\Delta\ell= -4$, having a final OAM index of $\ell_{\text{out}} = 2$ and a radial index of $p=2$. } 
  \label{fig:Conceptual figure}
\end{figure*}

To begin, let us suppose an LG mode is modulated with a phase profile, $\exp(i \Delta\ell \phi)$, where $\Delta\ell$ is an integer as shown in Fig. \ref{fig:Conceptual figure} (a) ($\Delta\ell =1$) and (b) ($\Delta\ell =2$) and $\phi$ is the azimuthal coordinate. 
Initially, the LG mode has the field function $\text{LG}_{\ell_{\text{in}}}(r,\phi) = f_{|\ell_{\text{in}}|} (r) \exp(i \ell_{\text{in}}\phi)$ where $\ell_{\text{in}}$ is the initial topological charge, and $f_{|\ell_{\text{in}}|} (r) =(\sqrt{2}r/w)^{|\ell_{\text{in}}|} \exp\left( -r^2/w^2 \right)$ is the radial envelope function for the LG mode, defined in cylindrical coordinates, $(r, \phi, z=0)$. We assume that the initial field does not have radial modes, i.e. initially $p = 0$. Here $w$ is the mode size of the embedded Gaussian envelope, where the Gaussian envelope is $\exp(-r^2/w^2)$. 
An example is illustrated in Fig. \ref{fig:Conceptual figure} (a), where a beam is numerically propagated for a mode with an initial OAM charge of $\ell_{\text{in}}= 2$. As the beam propagates, it acquires an additional OAM charge of $\Delta\ell = 1$, resulting in a final topological charge of $\ell_{\text{out}} = \ell_{\text{in}} + \Delta\ell = 3$. Notably, the beam in this example preserves its characteristic donut-like shape throughout propagation.  While the field initially does not have radial modes (nodes), it appears that in some cases they emerge as the beam propagates, as can be seen in the numerical propagation in Fig. \ref{fig:Conceptual figure} (b) for $\ell_{\text{in}} = -2 $ and $\Delta\ell = 2$ corresponding to an output  OAM topological charge of $\ell_{\text{out}} = 0$.
\begin{figure*}[t]
  \centering
  \includegraphics[width=0.9\textwidth ]{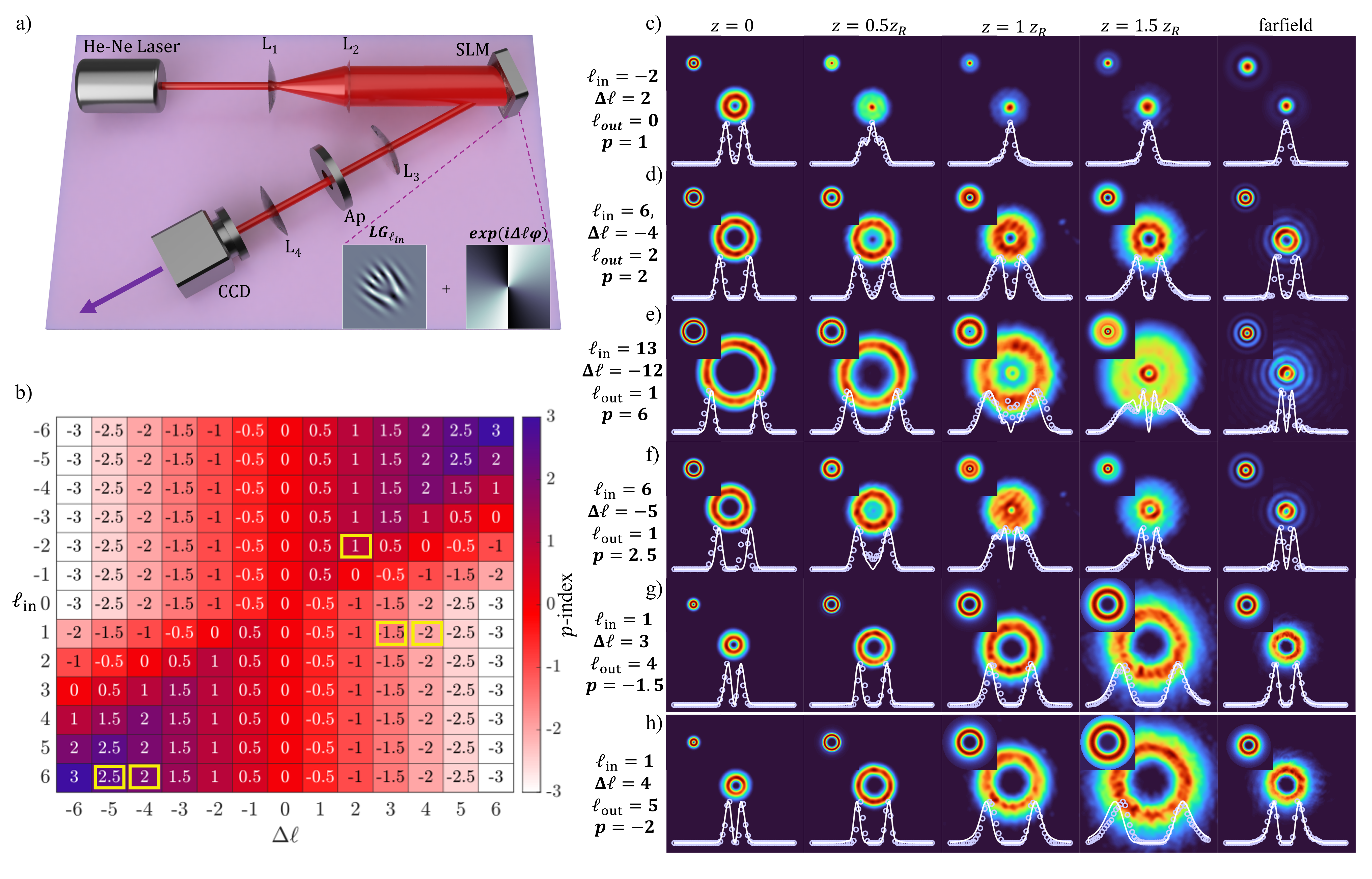}
  \caption[Experiment]{\textbf{Intensity profiles.} 
  \textbf{a)} Schematic of the experimental setup consisting of an expanded  laser beam (using lenses $L_{1(2)}$) that is incident on a spatial light modulator (SLM),  with the output mode imaged onto a camera (CCD) (using lenses $L_{3(4)}$). \textbf{b)} Heatmap of $p-$indexes (or radial index) associated with various choices of $\ell_{\text{in}}$ and $\Delta\ell$. The corresponding radial index is determined by $p = (|\ell_{\text{in}}| - |\ell_{\text{out}}|)/2$. Experimental, theoretical (top, left inset)  intensity profiles at propagation distances of $z = 0.0$, $z = 0.5 z_{R}$, $z = z_{R}$ and $z = 1.5 z_{R}$ and the far-field images for \textbf{c)} $\ell_{\text{in}}$ = -2 and $\Delta\ell$ = 2, \textbf{d)} $\ell_{\text{in}}$ = 6 and $\Delta\ell$ = -4, \textbf{e)} $\ell_{\text{in}}$ = 13 and $\Delta\ell$ = -12, \textbf{f)} $\ell_{\text{in}}$ = 6 and $\Delta\ell$ = -5, \textbf{g)} $\ell_{\text{in}}$ = 1 and $\Delta\ell$ = 3 and \textbf{h)} $\ell_{\text{in}}$ = 1 and $\Delta\ell$ = 4. The cross-sections are shown below each measured intensity plot, with the solid line corresponding to the theoretical intensities whereas the points correspond to the measured values.}
 \label{fig:Intensity profiles}
\end{figure*}
To compute the angular spectrum of the field, we assume Fraunhofer diffraction and exploit the cylindrical symmetry of the modes. Accordingly, the angular spectrum which describes the field can be computed from the Fourier transform
\begin{align}
\Psi_{\ell_{\text{out}}}(\textbf{k}) \propto \mathcal{F}(f_{|\ell_{\text{in}}|} (r) \text{e}^{i \ell_{\text{out}} \phi}) = \frac{1}{2\pi} \int f_{|\ell_{\text{in}}|} (r) \text{e}^{i \ell_{\text{out}} \phi} \text{e}^{i\mathbf{k} \cdot \mathbf{r}} d^2r,
\label{eq:farfieldmode1}
\end{align} mapping from position $\mathbf{r}= (r, \phi)$ to momentum space $\mathbf{k} = (k_r, \phi_k)$ where $k_r$ and $\phi_k$ are the radial and azimuthal polar coordinates in the momentum space, respectively. By applying the Jacobi-Anger expansion for plane-waves and solving the azimuthal and radial integral (that results in the $|\ell_{\text{in}}|$th order Hankel transform of the radial function $f_{|\ell_{\text{in}}|} (r)$), we arrive at the solution
\begin{align} 
\Psi_{\ell_{\text{out}}, p}(\textbf{k}) =& \mathcal{N}\ i^{-\ell_{\text{out}}} \text{e}^{i \ell_{\text{out}} \phi_k} \left( \frac{k_r w}{2} \right ) ^{\abs{\ell_{\text{out}}}} \text{e}^{-(k_r w/2)^2}   L_{ p}^{\abs{\ell_{\text{out}}}} \left[ \left( 
\frac{k_r w}{2} \right)^2 \right] ,
\label{eq:angularspectrumLG}
\end{align} where $\mathcal{N}$ is a normalization factor and the radial mode index is given by $p = \frac{1}{2}( |\ell_{\text{in}}|- |\ell_{\text{out}}|)$. To allow for half integer values for the radial index, we use the generalized Laguerre polynomials defined in terms of the confluent hypergeometric functions which can be defined as $L_p^{|\ell|}(z) = \frac{(|\ell| + 1)_p}{p!} \, {}_1F_1\left(-p; |\ell|+ 1; z\right)$, where $(\cdot)_{p}$ denotes the Pochhammer symbol. This allows the $p$-index to take negative values as well since it is possible to have  $|\ell_{\text{in}}| > |\ell_{\text{out}}|$.  The calculated  angular spectrum  that we have derived resembles that of eLG modes \cite{saghafi1998near} which have the characteristic form, $\text{eLG}^{m}_{p} (\rho, \phi_r) = \rho^{|m|}  \text{e}^{-\rho^2} L^{|m|}_{p}(\rho^2) \text{e}^{im \phi_r} $ for $\rho \equiv \frac{k_r w}{2}$ corresponding to the scaled momentum coordinates, confirming that these fields are indeed from the LG mode family. These modes are orthogonal in the OAM degree of freedom but not in the radial degree of freedom, and nonetheless, form an over-complete set. Figure  \ref{fig:Conceptual figure}
 (c)  shows exemplary analytical transverse angular spectral intensity profiles for several initial OAM charges ($\ell_{\text{in}}$) and transferred OAM charge pairs, i.e.  ($\ell_{\text{in}}$, $\Delta\ell$),  that in turn produce the total OAM charges ($\ell_{\text{out}}$) and radial index ($p$) pairs, i.e. ($\ell_{\text{out}}, p$).  For the first three cases, we see that the $p$-index satisfying $p\geq0$ corresponds to the number of dark rings that can be seen in the intensity profiles of the field. Whereas in the last two cases,  corresponding to $p<0$, we see that the field does not have dark rings even though the magnitude of the radial index is nonzero. 
To appreciate the significance of the radial index, we observe the behaviour of the angular spectrum, particularly the polynomial components. For $p \geq 0$, the generalized Laguerre polynomials, defined here in terms of confluent hypergeometric functions, remain polynomial in form. Consequently, we expect the radial profiles to exhibit nodes. This can be seen in the intensity and phase profiles for the first three examples corresponding to $p=$ 1, 2 and 6, of Fig. \ref{fig:Conceptual figure} (c) (see insets for phases). While the intensity does not have $p$ rings, the phase is seen to have $p+1$ sign changes in the radial direction.  That is to say, beginning at the center of the phase profile and moving radially outwards, there will be $p$ sudden phase changes. However, for the case where the modes have negative radial indices, shown in the last two examples of Fig. \ref{fig:Conceptual figure} (c), the modes are defined in terms of convergent series of Laguerre polynomials that have positive radial indices. In this case two types of solutions are expected; for integer negative values, the functions are still polynomials,  however, for non-integer negative values, the functions are infinite series of radial modes \cite{sephton2016revealing}. This results in a radial profile that still has an LG-like radial function, but without nodes ($p$-mode rings) in the radial direction, even through the index is non-zero. \\ 
 Nonetheless, the radial index we obtain resembles the version derived in Ref.~\cite{karimi2014radial} using the radial quantum operator which has states from the LG mode family as its eigenstates. Therefore this gives us an equation analogous to the principal quantum number, which here can be interpreted as the mode order following,
$|\ell_{\text{in}}| = 2p + |\ell_{\text{out}}|$, thereby setting constraints on the allowable radial indexes for a given initial ($\ell_{\text{in}}$) and output ($\ell_{\text{out}}$) OAM charge mapping. Moreover, this expression reflects the conservation of the total quanta in the initial field ($|\ell_{\text{in}}|$), suggesting that the transfer of orbital angular momentum (OAM), $ \Delta \ell = \ell_{\text{out}} - \ell_{\text{in}}$, is balanced by the introduction of radial modes to compensate for changes in the azimuthal component. \\
The far-field profile is thus given by
\begin{align} 
\Psi^{\text{Far}}_{\ell_{\text{out}}, p}(\textbf{k}, z)  =& \frac{\mathcal{N}} {\lambda z}  \text{e}^{i(kz - \pi/2 (\ell_{\text{out}} +1))}  \text{e}^{i\frac{z}{2k}k_r^2}  \text{e}^{i \ell_{\text{out}} \phi_k} \nonumber \\ & \times \ \  \left( \frac{k_r w}{2} \right ) ^{|\ell_{\text{out}}|} \text{e}^{-(k_r  w/2)^2} \ L_{p}^{
|\ell_{\text{out}}|} \left[ \left( \frac{k_r w}{2} \right)^2 \right] ,
\label{eq:farfieldLG}
\end{align}
and can be mapped to spatial coordinates through the expression $k_r = \frac{k}{z}r$ where k is the wavenumber. 
Further, the far-field expression includes the radial curvature ($\text{e}^{i\frac{z}{2k}k_r^2}$), the propagation phase $\text{e}^{ikz}$, and $i^{-\ell_{\text{out}}} = \text{exp}(-i\pi/2 \ell_{\text{out}}) )$ which we interpret to be associated with the Gouy phase  in the far-field. Accordingly, this result is consistent with the far-field mode function as derived in Ref. \cite{saghafi1998near} (See Equation (46) in the reference).

To estimate the  $z$-dependent function  we assume that it can be computed from  the Fresnel diffraction integral, $\psi_{\ell_{\text{out}}}(\textbf{r}) \propto \mathcal{F}(f_{|\ell_{\text{in}}|} (r) \text{e}^{i \ell_{\text{out}} \phi}\text{e}^{i\frac{k}{2z} r^2})$, which is similar to the previous Fourier integral but with the addition of a radially dependent phase modulation factor, $\text{e}^{i\frac{k}{2z} r^2}$. Solving this integral in cylindrical coordinates yields the expression,
\begin{align} 
\Psi_{\ell_{\text{out}}, p}(r, \phi, z)  &= \mathcal{N} \nonumber  \text{e}^{i(kz - \pi/2( \ell_{\text{out}} + 1))}     \left( \frac{q}{w(z)^2}\right) ^{1+p + |\ell_{\text{out}}|/2} \text{e}^ {i\frac{k  \ r^2}{2 \ R(z)}   } \\ &\times \text{e}^{i \ell_{\text{out}} \phi}    \left(\frac{\sqrt{q} r}{w(z)}  \right ) ^{|\ell_{\text{out}}|} \text{e}^{ - \frac{r^2 }{w(z)^2}}  L_{ p} ^{|\ell_{\text{out}}|} \left[     \frac{ q   r^2}{w(z)^2}   \right],
\label{eq:NearfieldLG}
\end{align}
where $w(z) = w_{0}\sqrt{1 + \left(\frac{z}{z_R}\right)^2}$, $q =  1+i \frac{z_R}{z} $,  $R(z) = (z^2 + z^2_R)/z$ is the radius of curvature,  $z_R = 1/2 k w^2$ is the Rayleigh range of the embedded Gaussian mode  from the initial beam LG${}_{\ell_{\text{in}}}(\cdot)$. Exemplary $z$-dependent intensity (Fig. \ref{fig:Conceptual figure} (d)) and phase (Fig. \ref{fig:Conceptual figure} (e)) profiles are shown using our analytical formulae in Eq. (\ref{eq:NearfieldLG}) (top panel)  and the numerically propagated simulations (bottom panel) using the angular spectrum propagation method. The analytical and numerical results are in perfect agreement producing a fidelity of $~100 \% $ in each case.
The experiment which proves our theory is seen in Fig. \ref{fig:Intensity profiles} (a). In this setup, a He-Ne laser beam ($\lambda$ =  633 nm) was passed through a 
lens system, consisting of lenses $L_{1(2)}$, before illuminating a phase-only spatial light modulator (SLM), so that the beam has constant intensity on the SLM. The SLM was then encoded with a complex amplitude hologram of an LG mode (of OAM, $\ell_{\text{in}}$, and $p=0$) that was multiplied by an additional azimuthal phase exp(i$\Delta\ell$) (see inset of Fig. \ref{fig:Intensity profiles} (a)). A 4f imaging system, consisting of lenses $L_3$ and $L_4$ 
allowed for spatial filtering of the desired, first diffraction order and for imaging the generated field on a Thorlabs CCD camera, 
which then captured intensity profiles at intervals from the image plane to 1.5 times the Rayleigh range and the far field. The combinations selected for $\ell_{\text{in}}$ and $\Delta\ell$ are highlighted (by yellow outlines) in the table of Fig. \ref{fig:Intensity profiles} (b), except for the case of Fig. \ref{fig:Intensity profiles} (e) which falls outside the plotted range of the heatmap. The intensity profiles obtained are presented in Fig.  \ref{fig:Intensity profiles} (c) to (h), where an initial OAM mode of $\ell_{\text{in}}$ is produced and then modulated with an azimuthal phase of $\Delta\ell$. Here the first three measurements, i.e. Fig. \ref{fig:Intensity profiles} (c-e) include cases for $(\ell_{\text{out}}, p )$ corresponding to (0, 1), (2, 2) and (1, 6) which were produced from pairs $(\ell_{\text{in}}, \Delta\ell)$ of (-2, 2), (6, -4) and (13, -12), respectively. These cases were chosen as they all produce positive radial indices.Next we show a result for a case where the radial index is a half-integer value, shown in Fig. \ref{fig:Intensity profiles} (f) for $(\ell_{\text{out}}, p) = (1, 2.5)$ which was produced by $(\ell_{\text{in}},  \Delta\ell)$ = (6, -5). While the cases shown so far correspond to positive integer values for the radial index ($p$), since $p=\frac{1}{2}(|\ell_{\text{in}}| - |\ell_{\text{out}}|)$, the $p$-index can be negative in value. We show measurements for such cases in Fig. \ref{fig:Intensity profiles}  (g) and (h) for $p = -1.5$ and $p=-2$,  for half-integer and integer cases,  with corresponding topological charges of $\ell_{\text{out}} = 4$ and $\ell_{\text{out}}=5$, respectively. These were produced for $(\ell_{\text{in}}, \Delta\ell)$ pairs corresponding to (1, 3) and (1, 4) for Fig. \ref{fig:Intensity profiles} (g) and (h), respectively. To measure the accuracy of the experimental results, a correlation between the experimental intensity profiles, and the theoretical (top, left inset) intensity profiles was performed for each of these mentioned cases. Each case produced correlations of above $90 \%$, showing a clear agreement, as can be confirmed by cross-sectional plots of the analytical (solid lines) and experimental results (points).\\
To conclude, we have examined the behavior of LG modes modulated by an azimuthal phase factor, and observed that the resultant modes have the angular spectrum of eLG beams, made apparent by their matching intensity profiles. Previously, it was deduced that modulation of LG modes with a phase-only spiral phase was deleterious as it reduced mode purity and beam power \cite{sephton2016revealing}, however, here we have shown that these modes have an analytical expression that can be interpreted again in terms of LG mode families (though it has been shown that they are also hypergeometric gaussian mode functions, \cite{karimi2007hypergeometric}). The link to eLG modes opens up the possibility of controlling the radial-index via phase-only control of the OAM content of an optical field; this unveils a coupling between the radial and azimuthal degrees of freedom. We emphasize that while the $p$-modes generated by these eLG-like fields are not orthogonal in the $p$-mode index, their ability to assume negative values is intriguing, as it arises from the conservation of the mode number. A detailed interpretation of this phenomenon will be explored in future studies.

\begin{backmatter}
\bmsection{Funding} The authors acknowledge financial support from Optica, the South African Quantum Initiative (SAQuTI) and the DSI Rentalpool programme.  

\bmsection{Disclosures} The authors declare no conflicts of interest.

\bmsection{Data Availability Statement} Data underlying the results presented in this paper are not publicly available at this time but may be obtained from the authors upon reasonable request.
\end{backmatter}




\end{document}